\documentclass[12pt]{article}
\pdfoutput=1

\usepackage{geometry}
\usepackage{amsmath}
\usepackage{amsfonts}
\usepackage{amssymb}
\usepackage{cite}
\usepackage{graphicx}

\usepackage{setspace}
\usepackage{hyperref}
\hypersetup {
	colorlinks=true,
	linkcolor=blue,
	linktocpage=true,
	citecolor=blue,
	urlcolor=blue
}

\newcommand{\be}{\begin{equation}}
\newcommand{\ee}{\end{equation}}
\newcommand{\ben}{\begin{eqnarray}\displaystyle}
\newcommand{\een}{\end{eqnarray}}
\newcommand{\mIm}{\text{Im}\,}
\newcommand{\mRe}{\text{Re}\,}

\numberwithin{figure}{section}

\numberwithin{equation}{section}

\date{}

\begin{document}
 \begin{center}
{
\Large{\bf High Energy Reflection Coefficients \\\vspace{5mm} of Black Holes and Branes}}

\vspace{10mm}

\textit{Roi Basha, Nissan Itzhaki and Lior Liram}
\break

Physics Department, Tel-Aviv University, \\
Ramat-Aviv, 69978, Israel \\

\end{center}

\vspace{10mm}
\begin{abstract}
We study the reflection coefficient associated with a wave that scatters off various  black holes in general relativity. We show that for large energies, compared to the scale set by the curvature, the reflection coefficient is suppressed exponentially with the energy. We find that the exponent  is fixed  by the singularity of the black hole to be $\frac12(\beta+\beta_{in})$ where $\beta$ is the inverse Hawking temperature and $\beta_{in}$ is the inverse temperature associated with the inner horizon. 
We also study the reflection coefficient associated with extremal D3-branes in string theory. In that case, the  exponent is determined by a fictitious singularity
located in the complexified space-time. Generalizations to M2-branes and M5-branes are also discussed. We  argue that $\frac12 \beta$ is a general bound on the exponent. 
\end{abstract}

\newpage
\tableofcontents

\section{Introduction}

The similarities between black hole (BH) physics and thermodynamics leads to several interesting exponential behaviors. The most known  is 
the thermodynamical nature of Hawking radiation \cite{Hawking:1974sw}. 

Another exponent is  the Lyapunov exponent that is related to the rate by which perturbations initially grow.  Only fairly recently \cite{Shenker:2013pqa, Maldacena:2015waa,Polchinski:2015cea} it was realized that on the BH side the Lyapunov exponent is determined by the shock wave interaction at the horizon that, to a large extent, is controlled by the exponential redshift
$e^{-2\pi t/\beta},$ where $\beta$ is the inverse Hawking temperature.

The eventual decay of the perturbation to a thermal state is also described by exponents that on the BH side are fixed by the quasinormal modes.   
For a Schwarzschild black hole the two exponents are of the same order, despite the fact that they are related to different regions in space-time. The shock wave interaction is related to the BH horizon and the quasinormal modes are associated with the entire region surrounding the BH.

The goal of the present paper is to present novel exponents associated with BHs and to pinpoint the regions in space-time that fix them.
We show that the reflection coefficient associated with a wave that scatters off the BH exhibits a   exponent 
\be\label{expC}
R\propto e^{-C E},
\ee
where $E$ is the energy of the wave that is taken to be large. Note that, unlike in the case of quasinormal modes (see e.g. \cite{Motl:2003cd}), we consider a  real (and positive) $E$.

We show that
\be\label{phy}
C=\frac12 (\beta+\beta_{in}),
\ee
where $\beta_{in}$ is the inverse temperature associated with the inner horizon. We work in conventions such that $\beta_{in}$ is negative. 

The reflection coefficient is determined by solving the wave equation outside the BH. Hence, one would expect $C$, much like the QNM, to be related to the region outside the BH. Surprisingly we find that it is fixed by the singularity.  
Causality implies that the region behind the horizon cannot affect the reflected wave. 
Still, as we show below, at high energies it is the singularity that determines the reflection coefficient in a rather interesting way. 

Scattering from BHs has been studied quite extensively in the past. Phase shifts and cross sections of fields of various spin (0, 1/2, 1, 2) scattering from various types BHs (Schwarzschild, Reissner-Nordstr{\" o}m, Kerr) were computed (See \cite{Futterman:1988ni} for references up to 1988. More recent work include \cite{Das:1996we,Doran:2005vm,Decanini:2011xi,Benone:2014qaa,Macedo:2013afa,Crispino:2009xt,Andersson:1994rk,Maldacena:1997ih,Glampedakis:2001cx,Dolan:2006vj}). 
Scattering from D-branes has also been studied (see, e.g., \cite{Klebanov:1995ni,Garousi:1996ad,Gubser:1996wt,Hashimoto:1996bf,Gubser:1997yh,Klebanov:1997kc,DAppollonio:2010krb,Bianchi:2011se}), mostly since they were identified with SUGRA's $p$-branes 
\cite{Horowitz:1991cd,Polchinski:1995mt}. In the present work, we study scattering of a massless scalar field from both BHs and branes. We emphasize the role of the singularity in determining the large energy reflection coefficient.

The paper is organized in the following way. In the next section we show that, at high energies, the reflection coefficient is fixed by the singularity for a general spherically symmetric BH. In section 3, we apply this for the Schwarzschild and Reissner-Nordstr{\" o}m BHs at $d\geq 4$ dimensions. In section 4, we demonstrate how our approach can extended to the D3 brane, as well as for the M2 \& M5 branes. In section 5 we argue that the  exponent is bounded from above by $\beta/2$. Section 6 is devoted to discussion.

\section{Generalities} \label{sec:generalities}

In this section we study general properties of  the reflection coefficient associated with probing the BH with a wave. We focus on the properties of the reflection coefficient at high energies compared to the energy scale set by the BH curvature, $1/M$, in the  Schwarzschild case.

In standard situations the reflection coefficient at  high energies is sensitive to the shape of the target at short distances. It is natural to expect that this is {\it not} the case when probing a BH at high energies. The BH singularity is shielded by the horizon. Causality, therefore, implies that the reflected wave cannot be sensitive to the BH singularity. The goal of this section is to show that this expectation is wrong and that there is a simple relationship between the BH singularity and the reflection coefficient at high energies. 

We start by illustrating how this surprising result comes about in a general spherical symmetric background in any dimension. Then, we demonstrate it in details for the 4D Schwarzschild black hole.   

\subsection{Scalar Fields in a Black Hole Background} \label{KG}

We consider a classical massless scalar field propagating in a $d$-dimensional background with metric $g_{\mu \nu}$. 
We focus on static spherically symmetric  black holes. For this type of backgrounds the metric takes the following form
\begin{align} \label{fmetric}
ds^2 = -f(r)dt^2 + f(r)^{-1}  dr^2 + r^2 d \Omega^2_{d-2}.
\end{align}
The zeros of $f(r)$ are the locations of the horizons and we also assume asymptotic flatness, i.e. $f(\infty)\to1$.

Our goal is to compute the reflection coefficient for a process in which  a wave arrives from infinity and is scattered from the black hole.
The field obeys the Klein-Gordon equation,
\begin{equation}\label{KG_equation}
\Box \phi = \frac{1}{\sqrt{-g} }  \partial_{\mu} (  \sqrt{-g}  g^{\mu \nu}   \partial_{\nu}  ) \phi =0.
\end{equation}
Instead of working with $r$ as the radial direction it is convenient to switch to the tortoise coordinate, defined by, 
\begin{equation} \label{ftort}
x(r) = \int \frac{dr}{f(r)},
\end{equation}
for which the (outermost) horizon is mapped\footnote{The horizon is mapped to $x \to - \infty$ whenever near the horizon $r_0$, we have $f(r)\sim (r-r_0)^a$ with $a\geq 2$. We assume this to be true.} to $x \to - \infty$ and infinity ($r\to \infty)$ is mapped to $x \to \infty$. 

After separation of variables, $\phi(t,x,\Omega) =r^{-\frac{d-2}{2}}(x) Y_{\ell m}(\Omega) e^{-i\omega t} \psi_{\omega \ell}(x)$,  the Klein-Gordon equation (\ref{KG_equation}) takes the form of a Schr{\"o}dinger-like equation for $x$,
\begin{align}\label{schrodingerEquation}
-\partial_{x}^2 \psi_{\omega \ell} + V \left(r(x)\right) \psi_{\omega \ell} =\omega^2 \psi_{\omega \ell}.
\end{align}
The potential is the Regge-Wheeler potential \cite{Regge} that reads, 
\begin{align}\label{ReggeWheeler}
 V (r) = f(r) \left( \frac{\ell \left(  \ell+d-3  \right)}{r^2} + \frac{\left( d-2 \right) \left( d-4 \right) }{4r^2} + \frac{ \left( d-2  \right)f'(r)}{2r}  \right).
\end{align}

Eq. \eqref{schrodingerEquation} suggests that the relation between $V(x)$ and the reflection coefficient, $R(\omega)$, is the standard one from quantum mechanics. In particular, since $V(x)$ vanishes both at the horizon ($x=-\infty$) and at infinity, at high energies we can use the Born approximation 
\begin{align}\label{BornApprox}
R(p)= \frac{1}{2ip}\int_{-\infty}^{\infty} e^{-2ip x}V(r(x))dx,
\end{align}
where $p$ is the momentum which is equal to the energy $\omega$ since we are considering a massless field.
This expression  supports the expectation mentioned above that the reflection coefficient is not sensitive to the BH singularity: the range of the integral corresponds to the region outside the BH as dictated by causality.

\begin{figure}
 \begin{center}
 \includegraphics[scale=0.5]{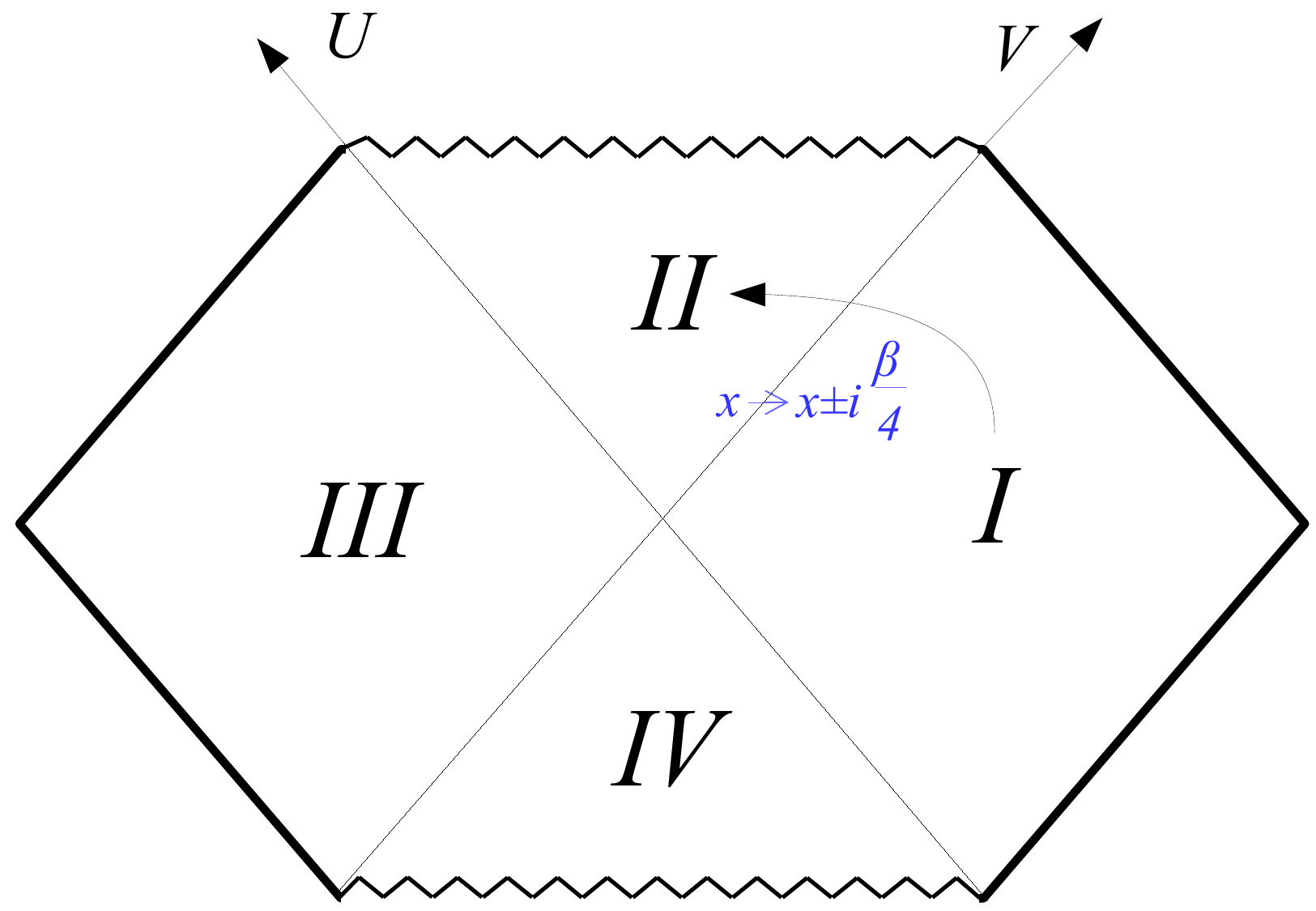}
  \caption{ Penrose diagram of the Schwarzschild BH showing the different regions. Moving between region amounts to shifting $x$ by an imaginary constant.}
  \label{SBHpen}
\end{center}

\end{figure}

There is, however, a hidden symmetry in this scattering problem which will allow us to relate the reflection coefficient to the singularity. It can be seen from the Kruskal coordinates, defined by 
\begin{align} \label{UV}
U = -e^{\frac{2\pi}{\beta}(x-t)},~~~ V = e^{\frac{2\pi}{\beta}(x+t)}.
\end{align}
As is well known, in these coordinates spacetime is extended to additional regions. As an example, the different regions of the Schwarzschild black hole are depicted in  figure \ref{SBHpen}.

The definition of the Kruskal coordinates, (\ref{UV}), implies that $U$ and $V$ are invariant under shift in the imaginary direction
\be
t\to t+\frac{i\beta}{2}(n+m),~~~x\to x+\frac{i\beta}{2}(n-m), ~~~\mbox{where}~~~n,m \in \mathbb{Z}.
\ee
In particular, keeping $t$ intact while shifting $x\to x\pm i\beta$ keeps us at the same point, say, in region I. Keeping $t$ intact while shifting $x\to x\pm i\beta/2$ takes a point in region I to its mirror point in region III. 
Since the potential in region III is identical to the potential in region I, we conclude that 
\be \label{periodv}
V(x)= V \left(x \pm i \beta/2 \right).
\ee

\begin{figure}
\centerline{\includegraphics[width=0.75\textwidth]{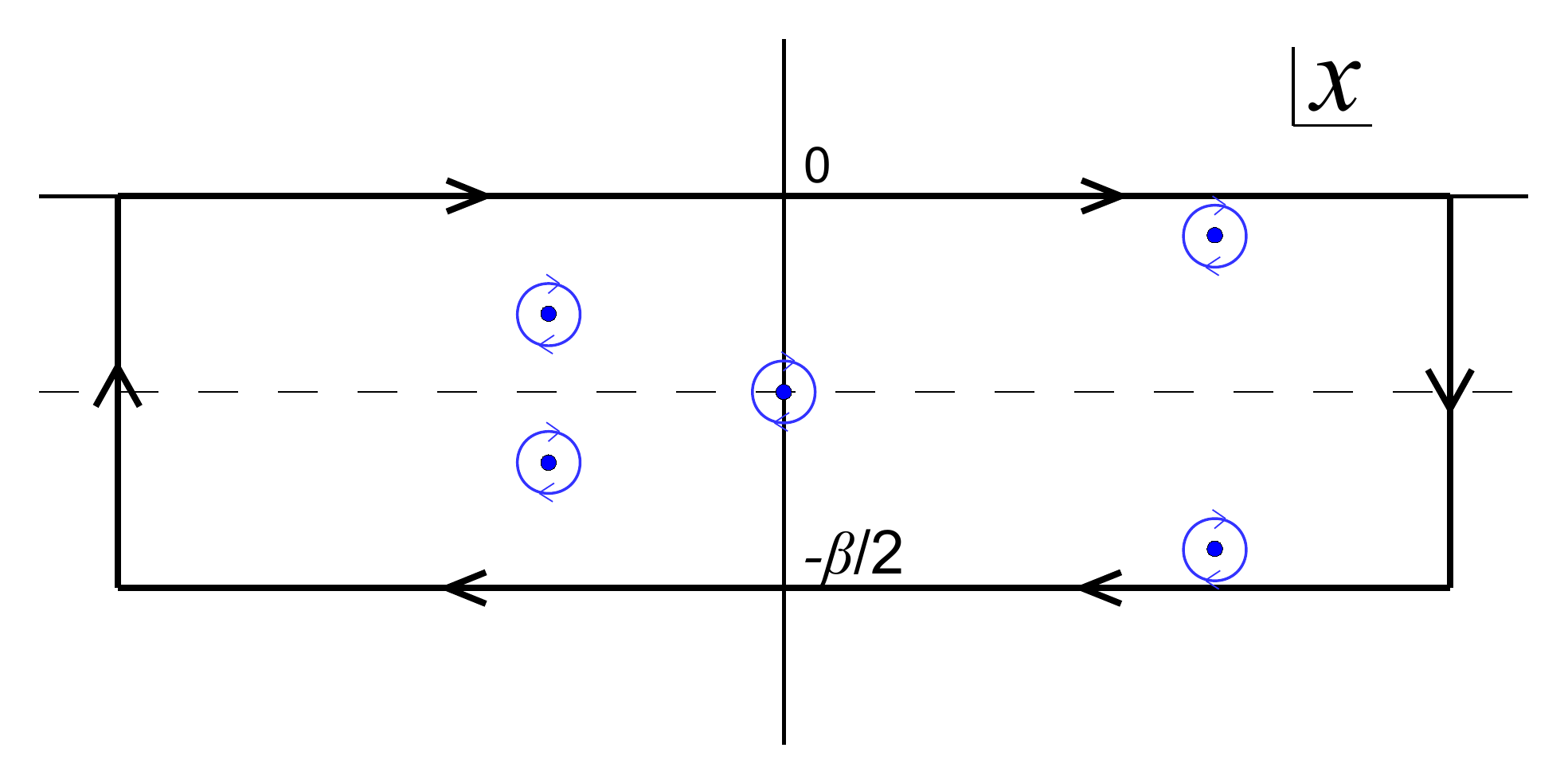}}
\caption{The contour used to calculate the reflection coefficient in the Born approximation. It goes clockwise around the strip $|\mRe{x}| \leq \infty,~ -\beta/2 \leq \mIm{x} \leq 0$. Blue dots represent possible locations of poles of $V(x)$. The large contour (black) is equivalent to the sum of the small contours (blue) encircling the poles.}
\label{fig:baseContour}
\end{figure}

To take advantage of this symmetry we evaluate the integral \eqref{BornApprox} in the complex plane.
We chose a rectangular contour, as plotted in figure \ref{fig:baseContour}. The two vertical lines do not contribute since the potential vanishes at $x = \pm \infty$. The contribution from the the two horizontal lines differ only by a factor of $e^{-\beta p}$, due to the symmetry \eqref{periodv}. Using the residue theorem we get,
\begin{align}\label{R_temp}
R(p) \left( 1-e^{-\beta p} \right)=-\frac{\pi}{p} \sum \text{Res}\left( V(x) e^{-2ipx} \right)+ \text{b.c.},
\end{align}
where the sum is over all the poles within the rectangle. `b.c.' stands for any contribution that may come from branch cuts. In the next subsection, we show that these contributions are always sub-leading at high energies.
 
In the high energy limit, $\beta p \gg 1$, we can ignore the factor of $e^{-\beta p}$ on the LHS of \eqref{R_temp} . Also, since there is an exponential suppression as we go down the imaginary axis, only the poles closest to the real line will contribute. We end up with,
\begin{align}\label{R_sum_of_poles}
R(p) =-\frac{\pi}{p} \sum_{\substack{\text{closest}\\ \text{poles}}} \text{Res}\left( V(x) e^{-2ipx} \right) +\text{b.c}.
\end{align}
This can be simplified even further. In appendix \ref{apc} we show that, to leading order, the potential always behave as $V \sim -\lambda(x-x_0)^{-2}$ near a singularity (pole) located at $x_0$. Therefore, we finally have,
\begin{align}\label{reflead}
R(p) = -2\pi i \sum_{i \,\in \,\substack{\text{closest}\\ \text{poles}}}\lambda_i  e^{-2ipx_i} +\text{b.c.},
\end{align}
where $\{x_i\}$ are the locations of the singularities. Eq. \eqref{reflead} shows quite nicely how, in contrary to what one might expect, the reflection coefficient at high energy depends only the behavior near the singularity. Technically speaking, this is the result of the symmetry \eqref{periodv} which allowed us to encircle the singularity in the complex plane. This can be viewed as a loophole in the argument that causality should prevent us from probing the singularity.

\subsection{4D Schwarzschild Black Hole and Branch Cuts}\label{sbh}
In this subsection we show how \eqref{reflead} applies for the case of a Schwarzschild BH in 4 dimensions. We also use this case to demonstrate that contributions from branch cuts are always sub-leading at high energies.

The metric of the Schwarzschild black hole in four dimensions is
\begin{align}
ds^2 = -f(r)dt^2 + f(r)^{-1}  dr^2 + r^2 d \Omega ^2  ~~~~\text{, with  }~~~~  f(r) = 1-\frac{1}{r} .
\end{align}
Here, for simplicity, we have set $2M=1$.  In these units the horizon is at $r=1$ and  the inverse temperature is $\beta = 8 \pi M = 4\pi$.
The tortoise coordinate is,
\begin{align}
 x =  \int f(r)^{-1}dr= r+ \log(r-1).
\end{align}

Focusing on the $\ell = 0$ case, the potential \eqref{ReggeWheeler} reads, 
\begin{equation}
V(r) = \left(	1-\frac{1}{r}	\right) r^{-3}.
\end{equation}
This potential has a pole only at the singularity ($r=0$). If we define $z(r) \equiv  x(r)-x_0$, where $x_0= -i \pi$ is the location of the singularity in the $x$-plane, then near the singularity the potential behaves as,
\begin{equation}\
V \sim -\frac{1}{4} z^{-2}.
\end{equation}
Plugging this into \eqref{reflead}, we obtain (after restoring the units),
\begin{align}  \label{sbhR}
R(p) \sim   e^{-\frac12 \beta p} = e^{-4\pi M p}. 
\end{align}

\begin{figure}
 \begin{center}
 \includegraphics[width=0.5\textwidth]{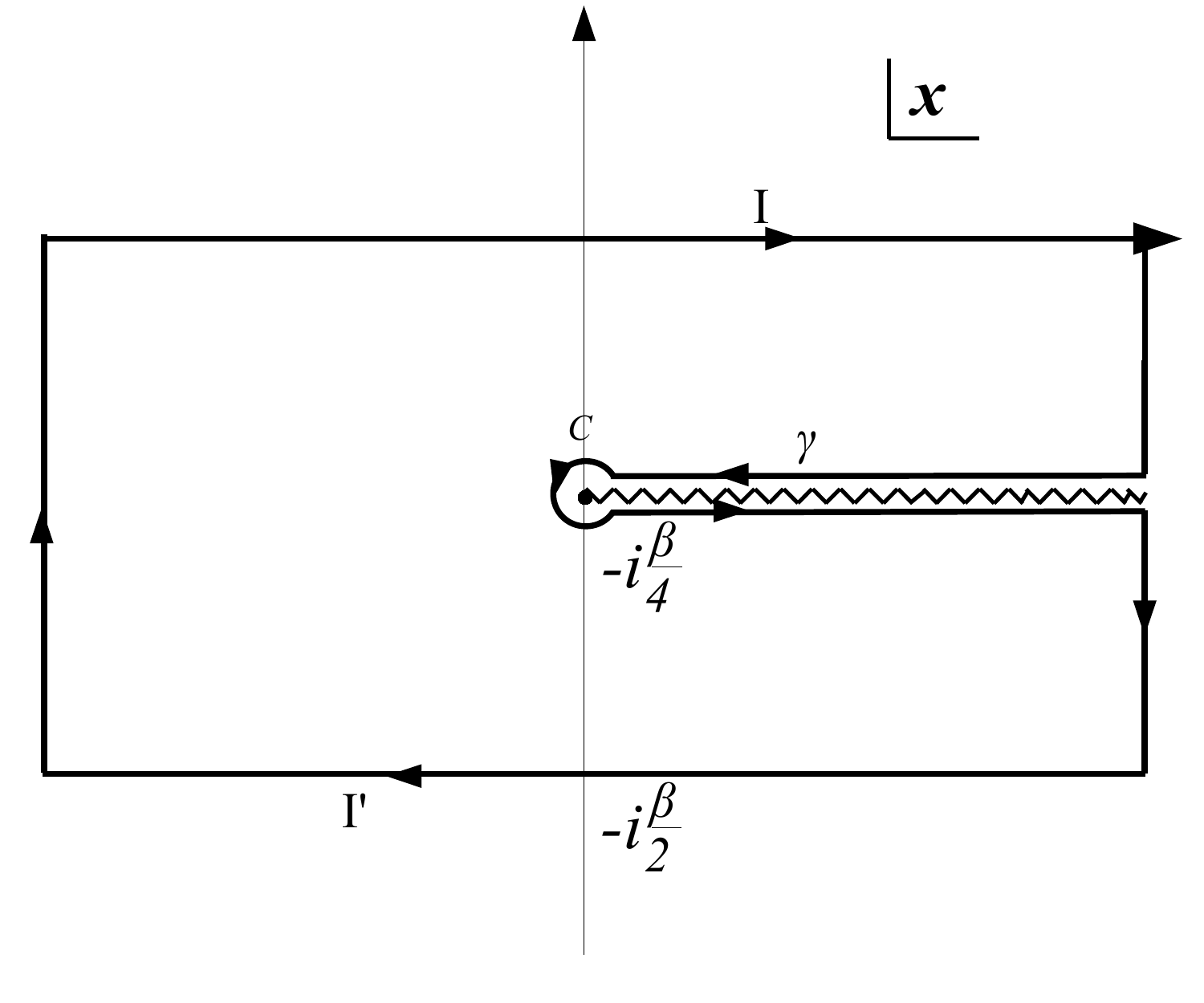}
  \caption{ The $x$-plane contour for the integral around the Schwarzschild singularity. A branch cut extends from the singularity to infinity and the contour is deformed accordingly.  }
\label{SBHcontour2}
\end{center}
\end{figure}

Finally, we wish to briefly discuss the issue of branch points. In the Schwarzschild BH (and in most other cases as well) the singularity is also a branch point of the potential. Including the sub-leading term as well, the potential near the singularity reads,
\begin{equation}\label{sbhPot_NLO}
V(z) \sim -\frac14 z^{-2} -\frac{i}{6\sqrt{2}}\,z^{-3/2} + \frac{1}{18}z^{-1} + \frac{\sqrt{2} i}{27} z^{-1/2}. 
\end{equation}
Therefore, the contour we have to take is not the one shown in figure \ref{fig:baseContour} but rather the one in figure \ref{SBHcontour2}, where we have chosen a branch cut that extends to infinity. Compared to figure \ref{fig:baseContour}, the new addition to the contour is the `keyhole' contour. Since $V(x)$ goes to zero at infinity, in order to evaluate the keyhole integral in the high energy limit, we only have to consider singular terms in the potential. Each singular term of the form, $z^{-q}$, contributes, 
\begin{equation}
 \int_{\text{keyhole}} \mkern-30mu e^{-2ipx} z^{-q}dx = c(q)\, p^{q-1} e^{-2 i p x_0},~~~\text{ with, }~~c(q)=\int_{\text{keyhole}} \mkern-30mu e^{-2iw} w^{-q}dw,
\end{equation}
so we only need to consider the most singular one which, as stated above, is always $z^{-2}$. Therefore, at leading order we may ignore any contributions that come from branch cuts and use only the first term in \eqref{reflead} to compute the reflection coefficient.

\section{Spherically Symmetric Black Holes}

In this section we use the method presented in the previous section to compute the high energy reflection coefficient for  spherically symmetric BHs in any dimension $d\geq 4$. The $d=4$ Schwarzschild case was already discussed in the previous section. We start by generalizing this result to $d>4$ and then turn to the electrically charged Reissner-Nordstr{\" o}m BH  in $d=4$ and its analogue in $d>4$. We show that in all cases the reflection coefficient is 
\begin{equation}\label{R_BH_gen}
R \sim e^{-\frac12 \left(\beta + \beta_{in}\right) \, p},
\end{equation}
where $\beta_{in}$ is an inverse temperature associated with the inner horizon. In particular, \eqref{R_BH_gen} is true  for the Schwarzschild BH where there is no inner horizon (i.e. $\beta_{in} \to 0$) and also for extremal BHs where both inverse temperatures diverge but their sum remains finite.

\subsection{$d > 4$ Schwarzschild Black Hole}

The first case we consider is a Schwarzschild black hole in $d>4$ dimensions. We will show that at high energies, the reflection coefficient behaves the same way as in the 4-dimensional case \eqref{sbhR}.

The Schwarzschild solution in $d>4$ is in the form \eqref{fmetric} with \cite{Myers:1986un}

\begin{equation} \label{f_Schwarz_d>4}
f(r)  = 1-\frac{2\mu}{r^{d-3}}, ~~\text{with}~~~~ \mu = \frac{8\pi M}{(d-2)A_{d-2}},
\end{equation}
where $A_{d-2}$ is the area of a unit $(d-2)$-sphere.
The horizon is at $(2\mu)^{1/(d-3)}$, but in fact there are other, complex, solutions to $f(r)=0$ giving us a total of $d-3$ ``horizons" at,
\begin{equation}
r_j  = (2\mu)^{1/(d-3)} \exp\left(\frac{2 \pi i j}{d-3}\right),
\end{equation}
where $r_0$ is the only physical one. To each such horizon we can associate an inverse temperature,
\begin{equation}
\beta_j = \frac{4\pi}{f'(r_j)} =  \frac{4\pi}{d-3} (2\mu)^{1/(d-3)}\exp\left(\frac{2 \pi i j}{d-3}\right),
\end{equation}
and the Hawking temperature is the one associated with the physical horizon, $\beta = \beta_0$. In terms of the $\beta$'s the tortoise coordinate has the following simple form,
\begin{equation}
x(r) =  r + \sum_{j=0}^{d-4} \frac{\beta_j}{4 \pi} \log \left(r-r_j\right).
\end{equation}

As we saw in the previous section, in order to compute the reflection coefficient at high energies we only need the behavior of the potential \eqref{ReggeWheeler} near the singularity. Similarly to the 4-dimensional case, in the $x$-plane, the singularity is mapped to 
\begin{equation}\label{singPos}
x_0 = - i \frac{\beta}{4} + C_R,
\end{equation}
where $C_R$ is some real constant.  Near the singularity the potential behaves as  $V(x) \sim -\frac14 (x-x_0)^{-2}$ and therefore, according to \eqref{reflead}, we again have\footnote{We comment that the real constant in \eqref{singPos} adds a factor of $\exp{\left(-2 i C_R \, p\right)}$ to $R(p)$. However, this has no physical significance since it can always be eliminated by a constant shift to $x$.},
\begin{equation}
R(p) \sim e^{-\frac12 \beta p}.
\end{equation}

To conclude, we see that even though the complex structure is different than the 4-dimensional case, we still end up with the same reflection coefficient. That said, the result still depends on the dimension through $\beta$.

\subsection{$d=4$ Reissner-Nordstr{\" o}m Black Hole}

Next, we look at the electrically charged Reissner-Nordstr{\" o}m (RN) BH. The metric is spherically symmetric and in the form \eqref{fmetric} with 
\begin{equation}\label{RN_f_of_r}
f(r) = 1-\frac{2M}{r} + \frac{Q^2}{r^2}.
\end{equation}
There are two solutions to $f(r) = 0$, giving two horizons,
\begin{equation}
r_\pm = M \pm \sqrt{M^2-Q^2},
\end{equation}
and we consider only the sub-extremal case, $Q<M$. As before, with each horizon we associate an inverse temperature,
\begin{equation}
\beta_\pm = \frac{4\pi}{f'(r_\pm)} = \pm \frac{4\pi r_{\pm}}{r_{+} - r_{-}},
\end{equation}
where the Hawking temperature is the one associated with the outer horizon, $\beta = \beta_+$. We work with the convention that $\beta_-$ is negative since this reflects the fact that at the inner horizon outgoing null trajectories converge rather than
diverge. 

\begin{figure}
\begin{center}
\includegraphics[width = 0.23 \paperwidth]{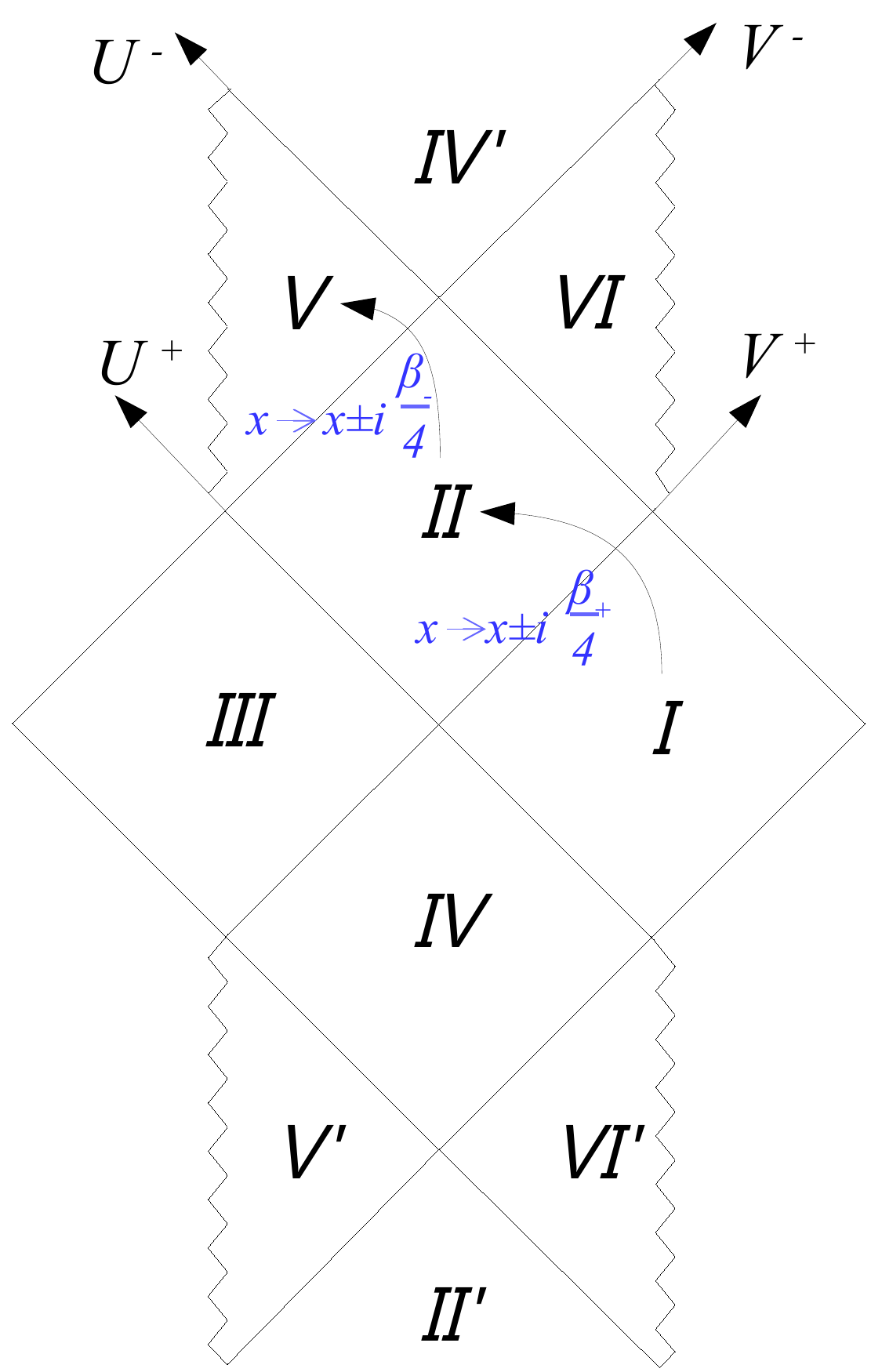}
\caption{ Penrose diagram of the RN Black Hole showing the different regions. Going between regions involves shifting the tortoise coordinate by an imaginary constant.}
\label{RNpenrose}
\end{center}
\end{figure}

The maximally extended manifold can be defined similarly to the Schwarzschild case but requires two sets of Kruskal coordinates,
\begin{align}
U^\pm = -e^{\frac{2\pi}{\beta_\pm}(x-t)} \text{~~and~~} V^\pm = e^{\frac{2\pi}{\beta_\pm}(x+t)},
\end{align}
where the tortoise coordinate is
\begin{equation}
x(r) = r +  \frac{\beta_+}{4\pi} \log\left(\frac{r}{r_+}-1\right)+ \frac{\beta_-}{4\pi} \log \left( \frac{r}{r_-}-1 \right)
\end{equation}
The Penrose diagram of the extended manifold is shown in figure \ref{RNpenrose}. The outer horizon corresponds to $U^+V^+=0$ and inner horizon to  $U^-V^-=0$. To go from region I to region II, one takes $x \rightarrow x \pm i \frac{\beta}{4} $ and $t \rightarrow t \mp i \frac{\beta}{4} $. Getting from region II to region V, is achieved by taking $x \rightarrow x \pm i \frac{\beta_-}{4}$, $t \rightarrow t \pm i \frac{\beta_-}{4}$. The singularity is, then,  mapped to
\begin{equation}
x_{\pm,n}=i \left(\frac{\beta_+ \pm \beta_-}{4} + \frac{\beta_+}{2}n\right); ~~~~ n \in \mathbb{Z}.
\end{equation}
where we have explicitly written the infinitely many copies due to the periodicity $x\to x + i \beta /2$.

\begin{figure}
 \begin{center}
 \includegraphics[scale=0.6]{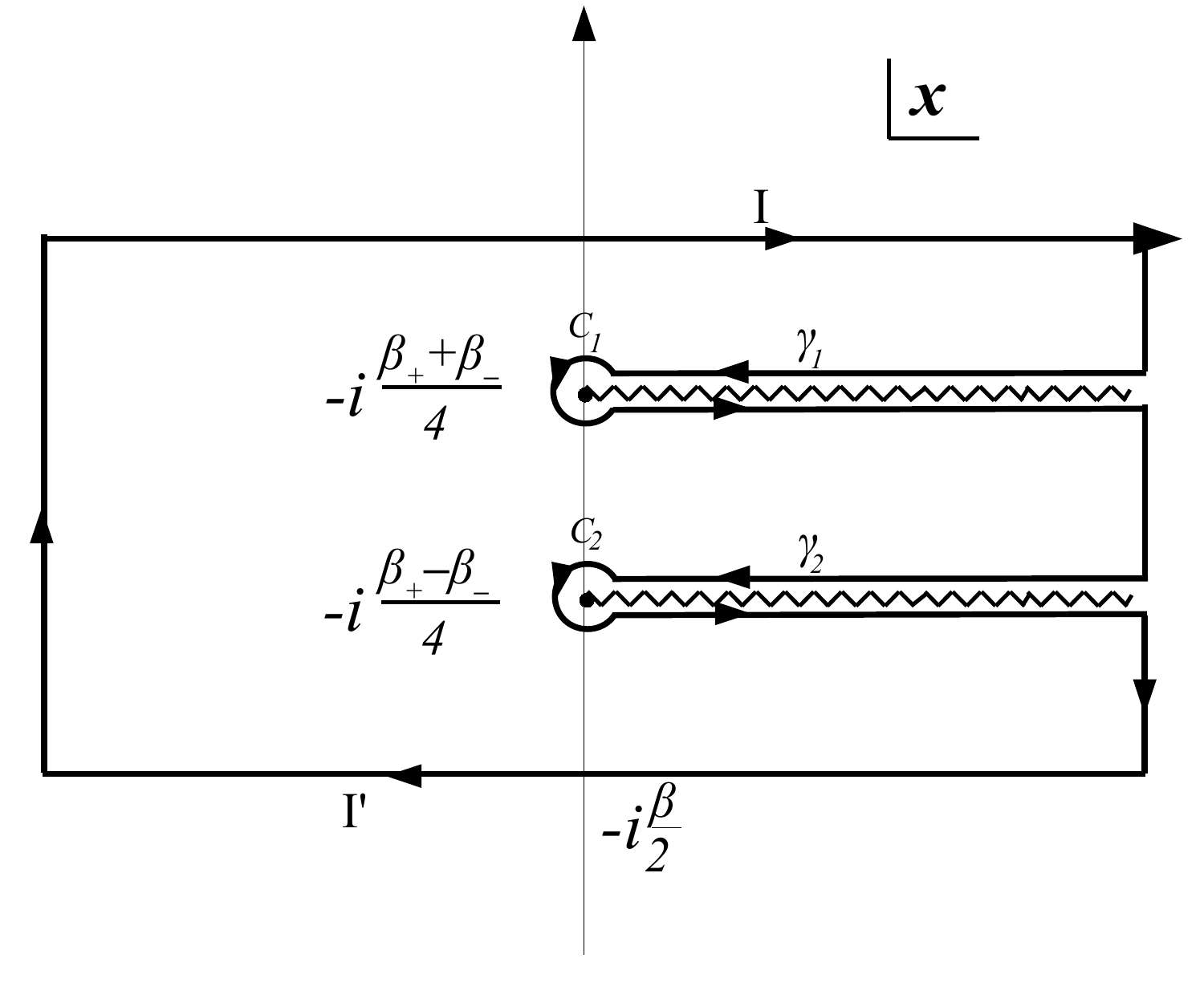}
  \caption{Contour used to compute $R(p)$ in the RN case. At high energy, the contribution from the branch cuts (jagged lines) is sub-leading leaving only the contribution from the poles at $-i (\beta_+ \pm \beta_-)/4$.}
\label{RN_contour}
\end{center}

\end{figure}

We will now compute the reflection coefficient in this background. The potential $V$ is obtained by plugging \eqref{RN_f_of_r} into \eqref{ReggeWheeler}. Near the singularities it behaves as,
\begin{equation}
V(x) \sim -\frac29 (x-x_\pm)^{-2},
\end{equation}
where we have omitted sub-leading terms that give rise to branch cuts. The contour we take to compute $R$ is plotted in figure \ref{RN_contour}. We see that both the singularities, $x_\pm$, lie within the strip\footnote{Note that $\beta_- <0$ and that $|\beta_-| < |\beta_+|$} $-\beta /2 <\mIm (x) < 0$. Yet, as discussed in section 3, the contribution from $x_-$, as well as the contributions from the branch cuts, are sub-leading. Thus, by using \eqref{reflead}, we get,
\begin{equation}\label{Ref_RN}
R(p) \sim e^{-\frac12(\beta_+ +\beta_-)p}.
\end{equation}
It is interesting that the  exponent depends on a property of the inner horizon. A similar feature was observed in \cite{Larsen:1997ge}. There it was shown that the area of the inner horizon, in certain cases, is related to dual quantities that can be defined outside the BH. 

Note that, since $\beta_+ +\beta_- = 8\pi M$, there is no dependence on $Q$ and we retain the result of the Schwarzschild BH \eqref{sbhR}.  This is not the case for $d>4$,  as we shall now see.

\subsection{$d>4$ Reissner-Nordstr{\" o}m Black Hole}

In this final subsection we consider the most general spherically symmetric BH in $d>4$ dimensions which is the analogue of the RN BH in higher dimensions. The metric is, again, in the form \eqref{fmetric} with \cite{Myers:1986un}
\begin{equation}
f(r) = 1-\frac{2\mu}{r^{d-3}} + \frac{q^2}{r^{2(d-3)}},
\end{equation}
where $\mu$ is given in \eqref{f_Schwarz_d>4} and  $q$ is related to the charge by,
\begin{equation}
q^2 = \frac{2}{(d-2)(d-3)}Q^2.
\end{equation}
Here too, we consider only the sub-extremal case, $q^2 < \mu ^2$.

Similarly to the $d>4$ Schwarzschild BH, there are multiple horizons located at,
\begin{equation}\label{HorizonsRN_d>4}
r_{\pm,n} = e^{\frac{2\pi i n}{d-3}} r_{\pm}, ~~~\text{with}~~~ r_{\pm} = \left( \mu \pm \sqrt{\mu^2-q^2} \right)^{\frac{1}{d-3}}.
\end{equation}
Only the two $n=0$ horizons are physical and are the analogues of the inner and outer horizons of the $d=4$ RN BH. Consequently, the maximally extended manifold also has the same global structure as in the $d=4$ case.

As we have done before, we associate a temperature with each horizon,
\begin{equation}
\beta_{\pm,n}  = \beta_\pm \exp{\left(\frac{2\pi i n}{d-3}\right)}, ~~\text{with}~~~ \beta_\pm = \pm \frac{4 \pi}{d-3} \,\frac{r_\pm^{d-2}}{r_+^{d-3}-r_-^{d-3}}
\end{equation}
and the Hawking temperature is that of the outer horizon, $\beta = \beta_+$. 
As we have done in the previous cases, we write the tortoise coordinate in terms of the $\beta$'s,
\begin{equation}
x(r) = r+ \sum_{\epsilon= \pm} \sum_{j=0}^{d-4} \frac{\beta_{\epsilon,j}}{4 \pi}  \log \left(  r-r_{\epsilon,j}  \right).
\end{equation}
The singularity is then mapped again to two points (plus copies due to the shift symmetry) in the $x$-plane,
\begin{equation}
x_{\pm}=i \frac{(\beta_+ \pm \beta_-)}{ 4}  + C_R,
\end{equation}
where $C_R$ is some real constant.

We see that the complex structure in the $x$-plane is identical to that of the 4-dimensional case. Therefore, to compute the reflection coefficient, we take the same contour used there (figure \ref{RN_contour}) and the leading contribution will again come from the upper pole. Near the singularity, the potential \eqref{ReggeWheeler} has, again, an inverse-square behavior,
\begin{equation}
V(x) \sim -\frac{(d-2)(3d-8)}{4 (2d-5)^2}(x-x_\pm)^{-2},
\end{equation}
so we use \eqref{reflead} to obtain,
\begin{equation}\label{Ref_RN_d>4}
R(p) \sim e^{-\frac12(\beta_+ +\beta_-)p}.
\end{equation}
This result looks identical to 4-dimensional one \eqref{Ref_RN} though it's quite different. In 4 dimensions the exponent does not depend on the charge since we had, $\beta_+ +\beta_- = 8\pi M$. This is no longer the case when $d>4$. Now we have,
\begin{equation}
\beta_+ +\beta_- =\frac{4\pi}{d-3} \frac{  \left(r_+^{d-2}-r_-^{d-2}\right)}{ \left(r_+^{d-3}-r_-^{d-3}\right)},
\end{equation}
and according to \eqref{HorizonsRN_d>4} this will be a function of $q$ for any $d>4$.

\subsubsection{The Extremal Limit}
We conclude this section with a short comment on the extremal limit, $q\to \mu$. In this limit, the inner and outer horizons coincide and the temperature goes to zero ($\beta \to \infty$). However,  \eqref{Ref_RN_d>4} is still valid since the sum of the two inverse temperatures is finite even though each of them diverges,
\begin{equation}\label{sumOfBetasExtremal}
\beta_+ +\beta_- \to 4\pi \frac{d-2}{(d-3)^2} \, \mu^\frac{1}{d-3}.
\end{equation}

This result can be verified directly as we now explain. In the extremal limit, the tortoise coordinate takes a slightly different form,
\begin{equation}\label{extremalTortoise}
x(r) = r -\frac{\mu\, r}{(d-3) \left(r^{d-3}-\mu\right)}+ \sum_{j=0}^{d-4} \frac{\tilde{\beta}_j}{4 \pi}  \log \left(  r-r_j  \right),
\end{equation} 
where $r_j =e^{\frac{2\pi i j}{d-3}} \mu^\frac{1}{d-3} $ are the horizons in the extremal case and
\begin{equation}\label{effectiveTemps}
\tilde{\beta}_j = 4\pi \frac{d-2}{(d-3)^2} \, \mu^\frac{1}{d-3} e^{\frac{2\pi i j}{d-3}},
\end{equation}
are \emph{effective} temperatures we associate with each horizon. The location of the singularity in the $x$-plane is determined by the effective temperature of the physical horizon,
\begin{equation}
\mIm (x_0) = -i \frac{\tilde{\beta}_0}{4},
\end{equation}
hence, according to \eqref{reflead} we have,
\begin{equation}
R(p) \sim e^{-\frac12 \tilde{\beta}_0 \, p}.
\end{equation}
Finally, comparing \eqref{sumOfBetasExtremal} and \eqref{effectiveTemps}, we see that we indeed have $\beta_+ +\beta_- \to \tilde{\beta}_0$.

\section{Brane Solutions}\label{sec:branes}
In this section we show that the method used in previous sections to calculate the high energy reflection can be applied to extended brane solutions as well.

Branes are quite different from the previously discussed cases. First, they are not spherically symmetric. Yet, there is a rotational symmetry in the directions transverse to the brane. The radial part of the Klein-Gordon equation can again be reduced to a Schr{\" o}dinger-like equation  where $\omega$ on the RHS of \eqref{schrodingerEquation} is replaced with $p_\perp$ (the momentum transverse to the brane). 

More importantly, we consider only \emph{non}-black (extremal) branes that have no singularities. Still, as we show below,  our approach is applicable since there are singularities in the complex $x$-plane. 

\subsection{D3 Branes}

The first case we discuss is that of a stack of $N$ coincident D3 branes. This background is given by \cite{Horowitz:1991cd},
\begin{equation}
ds^2= H^{-1/2}(r) (-dt^2 +d\vec y^{\,2}) + H^{1/2}(r) \left(  dr^2 + r^2 d\Omega^2_{5} \right),
\end{equation}
where $\vec{y} \in \mathbb{R}^3$  are the coordinates along the brane. The harmonic function is,
\begin{equation}\label{Dbrane_harmonicFunc}
H(r) = 1+\frac{L^4}{r^4}, ~~\text{with}  ~~L^4 = 4 \pi g_s N \alpha'^{2}. 
\end{equation}
In this background there is a horizon at $r=0$ and no singularities. However, in the complex $r$-plane there are $n$ `fictitious' singularities which are the roots of $H(r)$,
\begin{equation}\label{HarmonicRoots}
r_m = L \exp{ \left(\frac{ (2m+1)\pi i}{4}\right)}, ~~~ m=0,\dots,3.
\end{equation}
These fictitious singularities are a result of gluing the asymptotic region into $AdS_5\times S^5$. The celebrated low energy scattering on D3-branes \cite{Gubser:1996wt,Gubser:1997yh,Klebanov:1997kc} is not sensitive to these singularities, but they dominate the high energy scattering.

To see this, we compute the reflection coefficient following the same steps taken in previous sections. We start by switching to the tortoise coordinate, given here by,
\begin{equation}\label{brane_tortoise_2}
x(r) = \int  H^{1/2}(r) \, dr = r \,
   {}_2{\mkern -1mu}F_1\left(-\frac{1}{2},-\frac{1}{4};\frac{3}{4};-  \frac{L^4}{r^4}  \right).
\end{equation}
Just as we had for BHs, in terms of the tortoise coordinate the horizon ($r=0$) is mapped to $x\to -\infty$ and $r\to \infty$ is mapped to $x\to \infty$. In the $x$-plane, the singularities are located at,
\begin{equation}\label{D3_x_sings}
x_m = \zeta_4 r_m , 
\end{equation}
where we have defined,\footnote{This choice of notation is made clear in the following subsection.}
\begin{equation}
\zeta_{4} \equiv \, {}_2{\mkern -1mu}F_1\left(-\frac{1}{2},-\frac{1}{4};\frac{3}{4};1\right) = \frac{2 \sqrt{\pi } \,\Gamma \left(\frac{3}{4}\right)}{\Gamma \left(\frac{1}{4}\right)}.
\end{equation}

\begin{figure}[h]
 \begin{center}
 \includegraphics[scale=0.6]{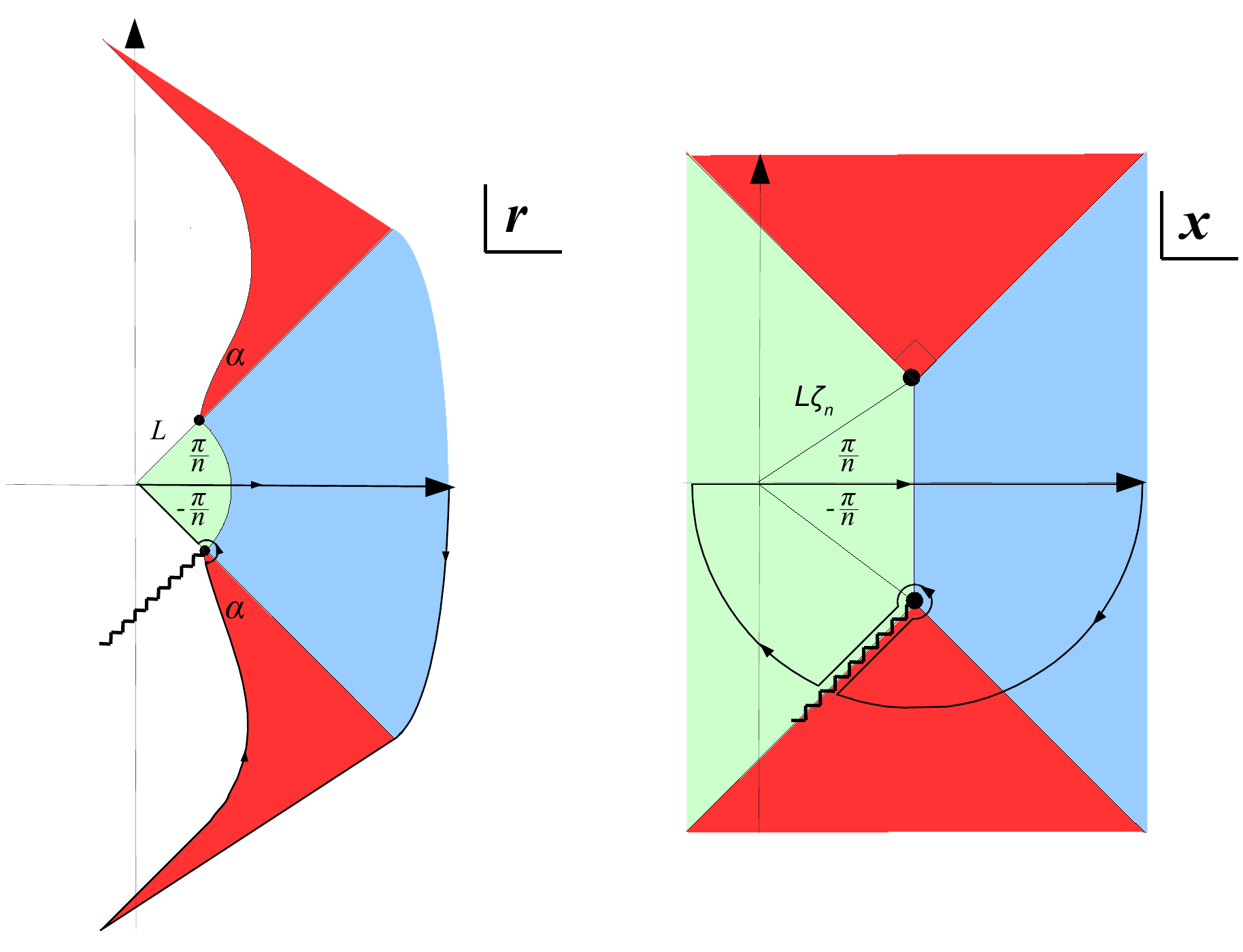}
  \caption{This plot shows how a patch in the $r$-plane (left) is mapped to the entire $x$-plane (right) for D-branes and M-branes. We have $n=4$ for D3 branes, $n=6$ for M2 and $n=3$ for M5. Also shown is the contour (in both $x$ and $r$ planes) on which the Born integral is evaluated.}
\label{branemaps2}
\end{center}
\end{figure}

According to \eqref{D3_x_sings} it appears as though there are 4 singularities in the $x$-plane, but that is misleading. The map $r \mapsto x$ \eqref{brane_tortoise_2} is not one-to-one and, in fact, it covers the entire $x$-plane multiple times. The branch we are interested in is the one that comes from a patch of the $r$-plane that contains the positive semi real axis (the physical region). This patch, shown in figure \ref{branemaps2}, contains only 2 singular points.

Next, we perform separation of variables to the Klein-Gordon equation to get,
\begin{equation}
\phi (t,x, \vec{y},\Omega) = r^{-5/2}(x) H^{-1/4}(r(x))e^{-i \left(\omega t - \vec{p}_y \cdot \vec{y}\right)} Y_{\ell m} (\Omega) \psi (x),
\end{equation}
where, as before, $\psi(x)$ obeys a Schr{\"o}dinger-like equation,
\begin{equation}
-\partial_{x}^2 \psi + V \left(r(x)\right) \psi = p_\perp^2 \psi, ~~ \text{with}~~~ p_\perp^2 = \omega^2 - \vec{p}_y^{\,2},
\end{equation}
the momentum in the directions transverse to the brane. The potential, in terms of $r$, reads,
\begin{equation}
V(r) = \frac{5 r^2 \left(3 r^4+L^4\right) \left(r^4+3 L^4\right)}{4\left(r^4+L^4\right)^3} + \frac{ \ell  (\ell +4)r^2}{r^4+L^4},
\end{equation}
and near the singularities it behaves as,
\begin{equation}\label{D3_pot_near_sing}
V(x) \sim -\frac{5}{36} (x- x_m)^{-2}.
\end{equation}
To compute the reflection coefficient we evaluate the Born integral \eqref{BornApprox} by closing the contour in the lower half plane\footnote{We note that as we had for BHs, the singularity is also a branch a point as could be seen by taking higher order terms in \eqref{D3_pot_near_sing}. Therefore, a branch cut is introduced and the contour is deformed accordingly. However, as discussed in previous sections, the contribution from the branch cut is sub-leading so we ignore it.}, as shown in figure \ref{branemaps2}. In the lower half-plane there is a single pole, $x_3 = \zeta_4 L \exp{\left(\frac74 \pi i\right)}$, and thus according to \eqref{reflead} we get,
\begin{equation}
R(p_\perp ) \sim e^{-\sqrt{2}\zeta_{4} L \,p_\perp},
\end{equation}
where we have omitted an unphysical phase, linear in $p_\perp$.

\subsection{M2 \& M5 Branes}

The next case we discuss is that of extremal M2 and M5 branes of 11-dimensional M-theory. This case is very similar to that of the D3 brane, as we shall now see. 

The metric of the M2 brane is \cite{Duff:1990xz},
\begin{gather}
ds^2_\text{M2} = H(r)^{-\frac{2}{3}} (-dt^2+d\vec y^{\,2}) + H(r)^\frac{1}{3} \left(  dr^2 + r^2 d\Omega^2_{7} \right),\\
\hspace{-144mm}\text{with,} \nonumber\\
H(r)  = 1+\frac{L^{6}}{r^{6}},  ~~ L^6 = 2^5 \pi^2 N \ell_p^6, \label{M2_harmonicFunc}
\end{gather}
and that of the M5 branes is \cite{Gueven:1992hh},
\begin{gather}
ds^2_\text{M5} = H(r)^{-\frac{1}{3}} (-dt^2+d\vec y^{\,2}) + H(r)^\frac{2}{3} \left(  dr^2 + r^2 d\Omega^2_{4} \right),\\
\hspace{-144mm}\text{with,} \nonumber\\
H(r)  = 1+\frac{L^{3}}{r^{3}},  ~~ L^3 = \pi N \ell_p^3,\label{M5_harmonicFunc}
\end{gather}
where $\ell_p$ is the 11-dimensional Planck length.

These are non-singular backgrounds, yet, as it happens for the D3 brane, the complex roots of $H(r)$ are singularities in the complex $r$-plane. The roots are given by,
\begin{equation}
r_m = L \exp{ \left(\frac{ (2m+1)\pi i}{n}\right)}, ~~~ m=0,\dots,n-1,
\end{equation}
where $n=6$ for M2 and $n=3$ for  M5.

To proceed, we need the tortoise coordinate. We notice that the $t,r$ components of the metric are not in the form \eqref{fmetric}, but rather in the form 
$
ds^2 = -a(r)dt^2 + b(r)^{-1}  dr^2 \,+  \dots.
$
In these cases, the tortoise coordinate is given by 
$
x = \int \left(a(r) b(r)\right)^{-1/2}dr.
$
Therefore, for M-branes, we get an expression similar to \eqref{brane_tortoise_2}, 
\begin{equation}
x(r) = \int  H^{1/2}(r)\, dr = r \,
   {}_2{\mkern -1mu}F_1\left(-\frac{1}{2},-\frac{1}{n};1-\frac{1}{n};-  \frac{L^{n}}{r^{n}}  \right).
\end{equation}
The singularities, $r_m$, are then mapped to 
\begin{equation}\label{brane_x_sings}
x_m = \zeta_n r_m,
\end{equation}
with,
\begin{equation}\label{zeta_m}
\zeta_n \equiv  \,  {}_2{\mkern -1mu}F_1\left(-\frac{1}{2},-\frac{1}{n};1-\frac{1}{n};1 
   \right) = \frac{\sqrt{\pi } \Gamma \left(\frac{n-1}{n}\right)}{2 \Gamma
   \left(\frac{3}{2}-\frac{1}{n}\right)}.
\end{equation}

From here everything follows through exactly as it did for the D3 brane. The Klein-Gordon equation is reduced to a Schr{\"o}dinger equation with a potential that behaves as \eqref{D3_pot_near_sing} near the singularities. Only one of which contributes to the Born integral. We end up with the following expressions
\begin{equation}
R_{_\text{M2}} (p_\perp)\sim e^{- L \zeta_{6} \, p_\perp} ~~\text{and}~~ R_{_\text{M5}}(p_\perp) \sim e^{-\sqrt{3} L \zeta_{3} \, p_\perp}
\end{equation} 
for the  reflection coefficients at high energies.

\section{A Bound on $R$}

In the previous sections we have shown that at high energies the reflection coefficient for BHs and branes is exponentially small, $R \sim e^{-C E}$. In this section  we show that the exponent, $C$, satisfies a bound,
\begin{equation}\label{theBound}
C \leq \frac{\beta}{2}.
\end{equation}
For all cases presented here this bound is satisfied: For extremal branes and BHs this is trivially true, as $\beta\to \infty$. For (non-extremal) BHs, we had $C = \frac12 (\beta + \beta_{in})$ where $\beta_{in} \leq 0$ and the Schwarzschild BH saturates the bound as $\beta_{in} \to 0$. Moreover, it was shown in \cite{Dijkgraaf:1991ba, Itzhaki:2018rld} that the $SL(2,\mathbb{R})/U(1)$ BH satisfies the bound, also when $\alpha'$ corrections are included. 

Let us  now show a general argument for this bound.
We consider systems that fall into the class described in section \ref{sec:generalities}. That is, configurations whose Klein-Gordon equation can be reduced to a Schr{\"o}dinger-like equation with a potential that satisfies the following properties:\\
\indent 1. It vanishes at $x\to\pm\infty$.\\ 
\indent 2. It has a shift symmetry  \eqref{periodv}. \\
\indent 3. It admits an inverse square behavior near the poles (see appendix \ref{apc}).\\
 As we have seen, the reflection coefficient for potentials with these properties is given by \eqref{reflead}. Therefore, a bound on $C$ is a bound on the position of the poles in the imaginary axis. Clearly, periodicity bounds the position of the poles by $-i\beta/2$ from below, giving  $C \leq \beta$. Yet, it is restricted even further by the following argument.

On the real axis the potential is real valued as it is a real function of the metric (and other fields such as  the dilaton when present). We may therefore apply the Schwarz reflection principle which states that in a connected domain, symmetric about the real axis and contains at least part of it, a function has the property,
\begin{equation}\label{SchwarzReflection}
V(\bar{x}) = \overline{V(x)},
\end{equation}
if it is holomorphic\footnote{The potentials we are considering are clearly \emph{not} holomorphic in the entire complex plane. They have singularities and branch cuts. We take a domain from which we cut out these non-holomorphic features. It is quite plausible that we can always chose branch cuts such that the remaining domain is connected.  } in the domain and real on the real axis. Combining \eqref{SchwarzReflection} together with the shift symmetry \eqref{periodv}, we find that
$V(x)$ has a reflection symmetry also across the line $\mIm(x) = -i \beta/4$, i.e.,
\begin{equation}
\overline{V\left(x-\frac{i\beta}{4}\right)}=V\left(\bar{x}-\frac{i\beta}{4}\right).
\end{equation}
This means that every pole in the strip $-\beta /2 <\mIm(x) <0$ has a `mirror' pole across the center of the strip (this is depicted in figure \ref{fig:baseContour}). Thus, the farthest a pole can be from the real line is $-i\beta/4$ where it coincides with its mirror image, giving the desired bound on $C$ \eqref{theBound}.

An interesting fact about the bound is that it depends only on the temperature. This may hint that the bound is not unique to gravity but in fact is more general. Perhaps this should not come as a surprise since it is merely the statement that BHs are nature's best absorbers. 

\begin{figure}
\centerline{\includegraphics[width=0.75\textwidth]{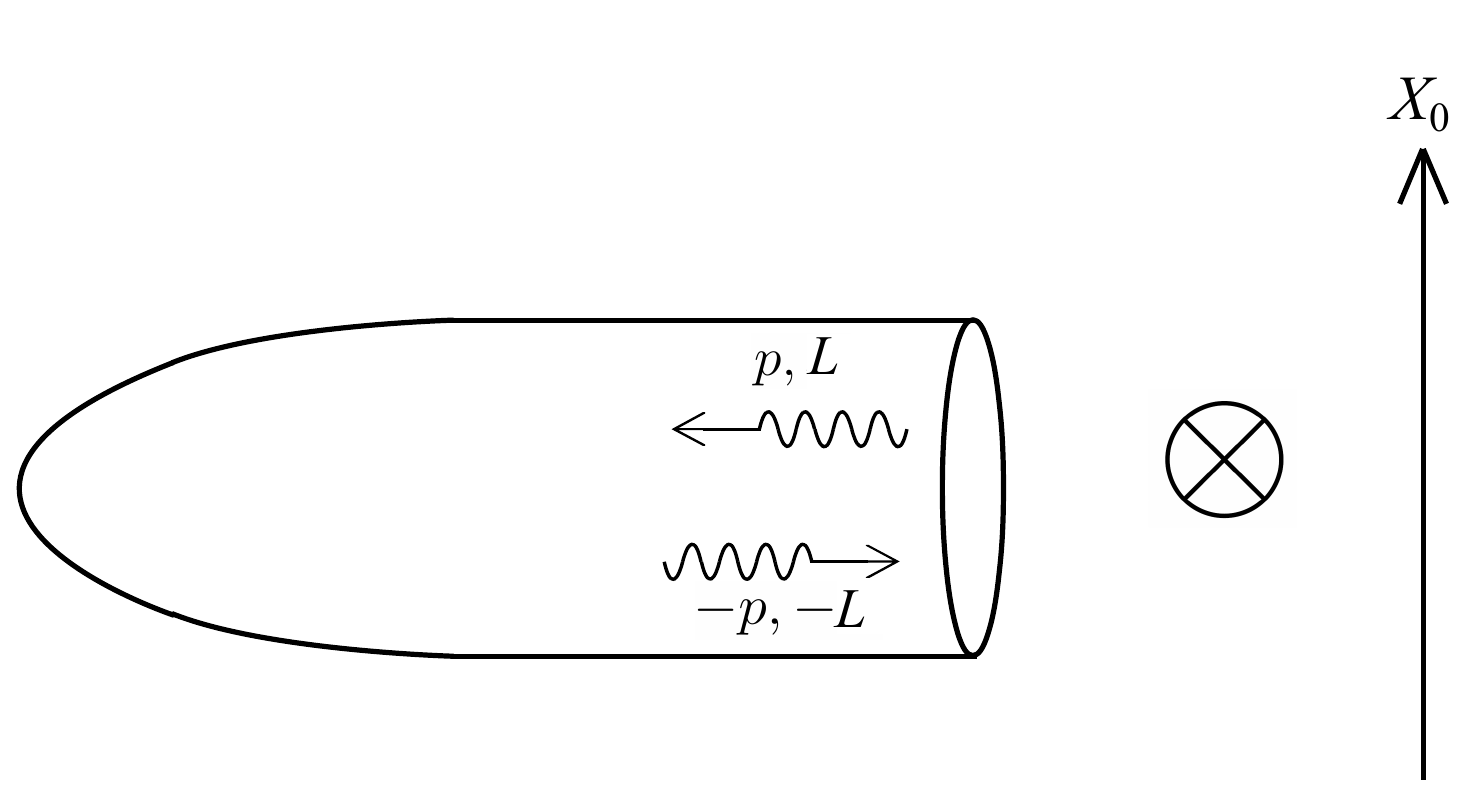}}
\caption{Scattering on the cigar: A wave with radial momentum, $p$, and angular momentum, $L$, is incident from infinity and reflected. An additional time direction $X_0$ was added to impose an on-shell condition. We note that at each point on the cigar there is a $(d-2)$-sphere (for a $d$-dimensional BH) that does not appear in the plot. }
\label{fig:cigarScatter}
\end{figure}

We end by pointing out a simple relation between the bound and scattering in the Euclidean setup. Consider a scattering process in the cigar geometry. A wave in the cigar in characterized by its quantum numbers at infinity: The radial momentum, $p$ and the angular momentum, $L$, associated with rotations around the axis of the cigar. To be on-shell we must add a time direction, $X_0$. So the setup is the one described in figure \ref{fig:cigarScatter}. 
In the eikonal limit, $p\to\infty$, the scattering reduces to the trivial phase shift associated with the approximated $\mathbb{R}^2$ at the tip of the cigar,
\be
R \sim (-)^L,
\ee
which upon Wick rotation, $L\to -i \frac{\beta}{2\pi} E$, saturates the bound.

Note that  $L$ and therefore also $E$ need not be large. In the Euclidean scattering the energy in the external time direction, $X_0$, is taken to be large (to satisfy the on-shell condition). After Wick rotation this means a  tachyon field with $m^2\to -\infty$. So from the BH point of view this is a highly non physical process. Still it is amusing that it saturates the bound for any $E$. 

\section{Discussion}
In this paper we illustrated that the reflection coefficient associated with scattering from a BH exhibits an  exponential suppression  at large energies. We calculated the  exponents in several cases. 

To a large extent, the  exponent is determined by the singularity. Quantum gravity corrections are expected to play an important role at the BH singularity. It is interesting that data that is accessible to an external observer  is, in principle,  sensitive to these effects. A concrete example of that are the perturbative and non-perturbative $\alpha'$ corrections that where considered in \cite{Ben-Israel:2017zyi,Itzhaki:2018rld}. 

\section*{Acknowledgments}

This work  is supported in part by the I-CORE Program of the Planning and Budgeting Committee and the Israel Science Foundation (Center No. 1937/12), and by a center of excellence supported by the Israel Science Foundation (grant number 1989/14). LL would like to thank the Alexander Zaks scholarship for providing
support of his PhD work and studies.

\appendix

\section{Behaviour of the Potential Near a Singularity} \label{apc}
In this appendix we show that, in almost all cases, the potential near a singularity admits an inverse square law behavior.
We consider a metric that is singular at some point $r=r_0$ and we also allow for a background dilaton field.
We write the metric and dilaton near the singularity,  to leading order in  $\epsilon = r-r_0$,
\begin{equation}
ds^2 \approx -\alpha_t \epsilon^{a}dt^2 + \alpha_r\epsilon^b d\epsilon^2 + \dots~~, ~~~
\sqrt{-g}e^{-2\Phi} \approx \gamma \epsilon^q,
\end{equation}
where the dots represent other coordinates (angular or otherwise) which are of no importance for our purposes. 
The tortoise coordinate near the singularity is then given by,
\begin{equation}\label{tortoiseNearSing}
x \approx \int  \sqrt \frac{\alpha_r \epsilon^b}{ \alpha_t \epsilon^a}  \ d\epsilon = x_0 + \sqrt\frac{\alpha_r }{ \alpha_t} \frac{2}{b-a+2} \epsilon^\frac{b-a+2}{2},
\end{equation}
where $x_0 \equiv x(r_0)$ is the location of the singularity in the $x$-plane. For consistency, we must assume\footnote{By making this assumption we exclude cases where the singularity occurs at infinite values of $x$.} that $b-a > -2$, so that the second term in \eqref{tortoiseNearSing} vanishes as $\epsilon \to 0$.

In this approximation, the Klein-Gordon equation becomes,
\begin{equation}
-\partial_x^2\phi - \frac{a + b - 2 q}{( a - b-2)}\frac{1}{x-x_0} \partial_x\phi=\omega^2 \phi,
\end{equation}
which is brought to a Schr{\"o}dinger form by setting, $\phi(x) = (x-x_0)^{\frac{a+b-2q}{4-2a+2b}} \,\psi(x)$. The potential turns out to be,
\begin{equation}\label{Vz}
V(x) \sim \left( \frac{(b-q+1)^2}{(b-a+2)^2} -\frac{1}{4} \right)~ (x-x_0)^{-2}.
\end{equation}
Note that  \eqref{Vz} is completely independent of the expansion coefficients: $\alpha_t, \alpha_r,\gamma$ -- the powers alone ($a,b,q$) determine the dominant part of the potential. Nevertheless, there could be cases where the coefficient in \eqref{Vz} vanishes and the leading behavior is obtained by taking higher order terms in the expansion. This happens when, $a+2q = 3b+4$ or $a+b=2 q$ (and $ b-q + 1 \neq 0$). This may indicate that $r_0$ is in fact a regular point or a horizon, since the potential has to be regular there. Yet, it may also happen for a singular point. We did not encounter such a case.

We conclude with a remark that inverse square law potentials, $V(x) = g x^{-2}$, were studied in the context of conformal quantum mechanics (see \cite{deAlfaro:1976vlx, Case}). For this type of potentials, unitarity restricts the coefficient to be $g\geq -1/4$. We see that \eqref{Vz} fulfills this requirement with the marginal value of $-1/4$ obtained for $q=b+1$.

\end{document}